\input{aipcheck}
\documentclass[
    ,final            
  ]
  {aipproc}

\layoutstyle{8x11double}

\begin{document}

\title{Hybrid $\gamma$ Doradus/$\delta$ Scuti Stars: Comparison Between Observations
and Theory}


\author{Bouabid, M.-P.}{
  address={UMR 6525 H. Fizeau, UNS, CNRS, OCA, Campus Valrose, 06108 Nice Cedex 2,
France}
  ,altaddress={Institut d'Astrophysique et de Géophysique de l'Université de Liège,
Allée du 6 Août, 17 4000 Liège, Belgium}
}

\author{Montalb\'an, J.}{
  address={Institut d'Astrophysique et de Géophysique de l'Université de Liège,
Allée du 6 Août, 17 4000 Liège, Belgium}
}

\author{Miglio, A.}{
  address={Institut d'Astrophysique et de Géophysique de l'Université de Liège,
Allée du 6 Août, 17 4000 Liège, Belgium}
}

\author{Dupret, M.-A.}{
  address={Institut d'Astrophysique et de Géophysique de l'Université de Liège,
Allée du 6 Août, 17 4000 Liège, Belgium}
}

\author{Grigahc\`ene, A.}{
  address={Centro de Astrofisica da Universidade do Porto, Rua das Estrelas,
4150-762 Porto, Portugal}
}

\author{Noels, A.}{
  address={Institut d'Astrophysique et de Géophysique de l'Université de Liège,
Allée du 6 Août, 17 4000 Liège, Belgium}
}

\begin{abstract}
$\gamma$~Doradus ($\gamma$~Dor) are F-type stars pulsating with  high order
$g$-modes. Their instability strip (IS) overlaps the red edge of the $\delta$~Scuti
($\delta$~Sct) one. This observation has led to search for objects in this region of
the HR diagram showing $p$ and $g$-modes simultaneously.
 The existence of such hybrid pulsators has recently been confirmed
\cite{handler2009} and the number of candidates is increasing (e.g.
\cite{matthews2007}). From a theoretical point of view, non-adiabatic computations
including a time-dependent treatment of convection (TDC) predict the existence of
$\gamma$~Dor/$\delta$~Sct hybrid pulsators (\cite{dupretgrigahcenegarrido2004},
\cite{grigahcenemartinruizdupret2006}).
Our aim is to confront the properties of the observed hybrid candidates with the
theoretical predictions from non-adiabatic computations of non-radial pulsations
including the convection-pulsation interaction.
\end{abstract}

\maketitle


\section{$\gamma$~Dor/$\delta$~Sct hybrid candidates}

\begin{figure}
  \includegraphics[height=.3\textheight]{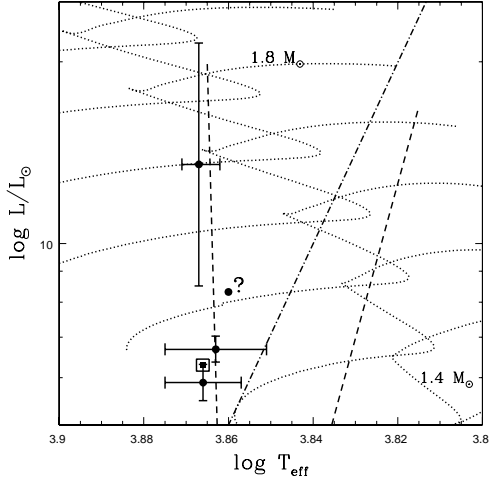}
  \caption{HR diagram location of the four hybrid pulsator candidates  (from bottom
to top: HD~8801, HD~49434, BD+18~4914, HD~114839) and their 1~$\sigma$ error
boxes, except for BD+18~4914 whose $T_{\rm eff}$ and $L$ uncertainties were not
available. Dashed lines represent the observational $\gamma$~Dor IS
\cite{handlershobbrook2002}, the dotted-dashed~line~represents~the~red~edge~of~the~$\delta$~Sct~IS~\cite{rodriguezbreger2001}.
The empty square locates the model we selected for our non-adiabatic
analysis. Evolutionary tracks computed with $X=0.7$, $Z=0.02$, $\alpha_{\rm MLT}=2.0$ and
$\alpha_{\rm ov}= 0.0$ are shown in dotted lines.}
  \label{HRdiagram}
\end{figure}

There are presently three  $\gamma$~Dor/$\delta$~Scuti hybrid pulsator candidates,
HD~49434 \cite{uytterhoevenmathiasporetti2008}, 
HD~114839 \cite{kingmatthewsrowe2006} and BD+18~4914
\cite{rowematthewscameron2006}. One more object, HD~8801 was already
proposed as
$\gamma$~Dor/$\delta$~Scuti pulsator by \cite{henryfekel2005} and has recently been confirmed as
a hybrid pulsator  \cite{handler2009}.
The available stellar parameters for these four stars have been collected from
literature and summarized in Table~1.
In Figure~\ref{HRdiagram} we plot their location in the HR diagram as well as the
observational $\gamma$~Dor instability strip \cite{handlershobbrook2002} and the
red edge of the $\delta$~Sct instability domain \cite{rodriguezbreger2001}. We note that  these
four stars have quite close $T_{\rm eff}$ (within 100~K),
 and are located near the blue edge of the $\gamma$~Dor IS and inside the 
$\delta$~Sct IS.

\begin{table}
\begin{tabular}{lcc}
\hline
&\bf HD 8801                                         & \bf HD 49434    \\ \hline
Spectral type                           & A7m             & F1V   
            \\
Parallax $\pi$ (mas)                    & 17.91 $\pm$ 0.75$_{(18)}$& 24.94 $\pm$
0.75$_{(18)}$  \\
$T_{eff}$ (K)                           & 7345 $\pm$ 155$_{(10)}$  & 7300 $\pm$ 200$_{(23)}$   \\
$\log g$ (cgs)                          & 4.2$_{(12)}$ & 4.4 $\pm$ 0.2$_{(23)}$   \\
$\log \left(\frac{L}{L_{\odot}}\right)$ & 0.77 $\pm$ 0.03$_{(10)}$ & 0.825 $\pm$
0.022$_{(23)}$ \\
$[Fe/H]$ (dex)                          & -                & 0.10 $\pm$ 0.12$_{(23)}$        \\
$R$ ($R_{\odot}$)                           & 1.7 $\pm$ 0.1$_{(12)}$   & 1.60 $\pm$
0.05$_{(16)}$   \\
$v\,\sin\,i$ (km~s$^{-1}$)                & 55 $\pm$ 5$_{(12)}$    & 87 $\pm$ 4$_{(23)}$    \\
$M$ ($M_{\odot}$)                           & 1.54 $\pm$ 0.03$_{(10)}$ & 1.55 $\pm$
0.14$_{(2)}$  \\ \hline
& \bf HD 114839         & \bf BD+18 4914 \\ \hline
Spectral type                           & Am & F5
(Am?$_{(16)}$)   \\
Parallax $\pi$ (mas)                    & 5.04 $\pm$ 1.04$_{(18)}$ & -             
          \\
$T_{eff}$ (K)                           & 7356 $\pm$ 77$_{(16)}$ & 7250$_{(19)}$   \\
$\log g$ (cgs)                          & 4.39 $\pm$ 0.5$_{(1)}$ & 3.77$_{(19)}$   \\
$\log \left(\frac{L}{L_{\odot}}\right)$ & 1.132 $\pm$ 0.18$_{(S)}$ & 0.92$_{(19)}$   \\
$[Fe/H]$ (dex)                          & 0.04 $\pm$ 0.15$_{(1)}$ & -             
          \\
$R$ ($R_{\odot}$)                           & 2.177 $\pm$ 0.450$_{(16)}$ & -        
               \\
$v\,\sin\,i$ (km~s$^{-1}$)                & 66.7 $\pm$ 5.0$_{(1)}$ & -             
          \\
$M$ ($M_{\odot}$)                           & -                       & -           
            \\
\hline
\end{tabular}
\caption{Stellar parameters of the four hybrid $\gamma$~Dor/$\delta$~Sct candidates.}
\label{hybridparam}
\end{table}

\section{Theory versus observations}

\begin{figure}
  \includegraphics[height=.3\textheight]{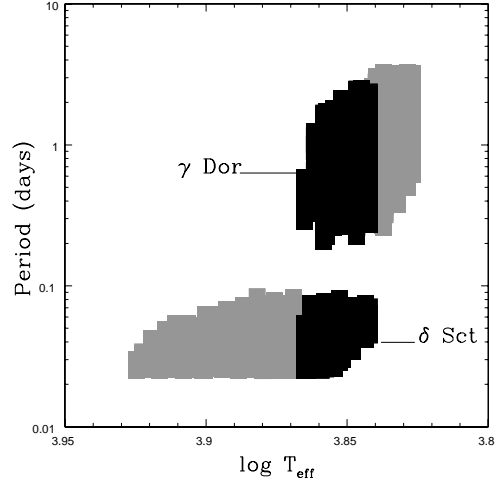}
  \caption{$\log T_{\rm eff}$--$\log P$ theoretical instability domain for $\ell=0$--4 modes
  for $\gamma$~Dor and $\delta$~Sct models with $X=0.7$, $Z=0.02$, $\alpha_{\rm MLT}=2$, $\alpha_{\rm ov}=0$ and $1.2$ $M_\odot$ $<M<2.5$ $M_\odot$.~In
  grey are the $\gamma$~Dor and $\delta$~Sct domains and in black is the hybrid~domain.}
  \label{pteff}
\end{figure}

To study the pulsation properties of these stars we have at our disposal a grid of stellar models
computed with the evolution code CL\'ES \cite{scuflairemontalbantheado2008}. The grid
properties are the following: stellar masses range  from 1.2 to 2.5
$M_{\odot}$ with a step of 0.1; four different chemical compositions described by the
metal mass fraction $Z=0.01$ and 0.02 with a hydrogen mass fraction $X=0.70$ and 0.73 are available. Moreover
three different values of the mixing length parameter of convection
($\alpha_{\rm MLT}=1.4$, 1.7, 2.0) and two values of the overshooting parameter ($\alpha_{\rm ov}= 0.0$, 0.2) 
can be chosen. The pulsation analysis is
done by using a version of the non-adiabatic pulsation code MAD  that includes the
effects of the  convection-pulsation interaction (\cite{dupret2001}, \cite{grigahcenedupretgabriel2005}). In fact, it is necessary to include the
effect of convection in order to match the observational red edge of the
$\delta$~Sct IS and  therefore to  study the hybrid pulsators.  Dupret et al. \cite{dupretgrigahcenegarrido2004} found
that a value of $\alpha_{\rm MLT}=2.0$ is necessary to fit the location of observational $\gamma$~Dor and
$\delta$~Sct IS (see also \cite{houdek2000}). In this preliminary analysis, we have restricted our choice of parameters
to $X=0.70$, $Z=0.02$, $\alpha_{\rm MLT}=2.0$ and $\alpha_{\rm ov}=0.0$.
The theoretical instability domain is shown in Figure~\ref{pteff}.

We computed an additional  main-sequence model (thereafter H model) whose 
HR diagram location is close to that of HD~49434 and HD~8801 (square symbol
in Figure~\ref{HRdiagram}). The parameters of that model are given in
Table~\ref{modelparam}. 

\begin{table}
\begin{tabular}{lr|lr}
\hline
$M$ ($M_{\odot}$) & 1.54 & $\alpha_{\rm MLT}$  & 2.0 \\
$R$ ($R_{\odot}$) & 1.552 & $\alpha_{\rm ov}$  & 0.0 \\
$T_{eff}$ (K) & 7346 & $X$ & 0.70\\
$\log g$ (cgs) & 4.24 & $Z$ & 0.02 \\
$\log \left(\frac{L}{L_{\odot}}\right)$ & 0.799 & & \\
\hline
\end{tabular}
\caption{Parameters of the H model}
\label{modelparam}
\end{table}

\begin{figure}
  \includegraphics[height=.39\textheight]{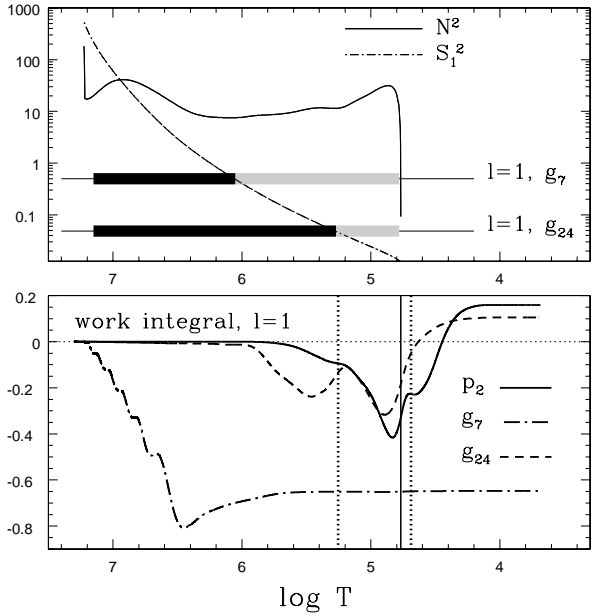}
  \caption{Propagation diagram (top) $\&$ work integrals (bottom) for $\ell=1$ $g_{24}$, $g_{7}$ and $p_{2}$ 
  modes of the H model. Top panel: thick dark lines: propagation regions, thick grey lines:
  evanescent regions for $g$-modes.
  Bottom panel: vert. continuous line:
  base of the CE; vert. dotted lines: Fe opacity bump ($\log T\,\sim\,5.3$) $\&$ He partial ionization zone ($\log T \,\sim\,4.9$).}
  \label{work}
\end{figure}

The pulsation analysis of the H model for  mode degrees $\ell=0$--8 reveals
that it behaves as a hybrid pulsator, with unstable $\gamma$~Dor modes
(from~$\nu=$~0.789~c/d~to 7.03 c/d) as well as unstable $\delta$~Sct modes (from $\nu=$ 11.0~c/d
to  49.0~c/d) separated by a region of stable modes. 
 To illustrate that we chose three different modes: a typical $\gamma$~Dor
$g$-mode ($\ell$=1, $g_{24}$), a typical $\delta$~Sct $p$-mode ($\ell$=1, $p_2$)
and a stable mode between $\gamma$~Dor and $\delta$~Sct frequency ranges ($\ell$=1,
$g_7$). In Figure~\ref{work} we present their propagation diagram and
their work integrals. Regions where the work
increases (vs. decreases) are driving (vs. damping) the oscillations. For the
$\gamma$~Dor $g$-mode, we see a clear driving mechanism at the location of the Fe
opacity bump ($\log T \,\sim\, 5.3$) but this $\kappa$-mechanism is not sufficient
to globally excite the mode. The main driving occurs at the base of the
convective envelope (CE) by the flux blocking mechanism (FBM) (in agreement with  \cite{guzikkayebradley2000}
and \cite{dupretgrigahcenegarrido2004}). For the $p$-mode there
is a contribution of the FBM at the base of the CE
and a contribution from the $\kappa$-mechanism in the He partial
ionization zone.
For the stable mode, there is an efficient radiative damping in the inner layers of
the star. Furthermore, the amplitude of the eigenfunction in the outer layers is small due to the presence of a large 
evanescent region before reaching the base of the CE, which inhibits the FBM (see also \cite{dupretmigliomontalban2008}).

\subsection{Rotational splitting: application to HD~8801 and HD~49434}

Even if theory predicts hybrid pulsators, we should wonder if the high frequency
modes detected in HD~8801 and HD~49434 are really $\delta$~Sct modes or
prograde $g$-modes moved to higher frequencies due to rotational splitting. 
It is well known that for modes with a pulsation frequency (PF) $\sigma$  comparable to,
or lower than, the rotational frequency $\Omega$, the Coriolis force 
term plays a major role in the equation of motion and the perturbative approach  is
no longer valid. Since  $\gamma$~Doradus stars show low-frequency $g$-modes, the
effects of the Coriolis force cannot be neglected  even if the star does not rotate
fast. Dintrans $\&$ Rieutord \cite{dintransrieutord2000} showed  that the
perturbative treatment of rotation is no longer valid for $\gamma$~Dor
with  rotation period smaller than $\approx 3$ days. Moreover, one should also
take into account the effect of rotation on the mode excitation
\cite{saiocameronkuschnig2007}. Nevertheless, 
in a first approximation, we estimate the rotational splitting by using the
perturbative approach at first order:
$$\sigma_{\rm obs}=\sigma_0+m\,\beta\,\Omega$$
with $\sigma_{\rm obs}$ the PF in the observer frame, $\sigma_0$
the PF in the corotating frame, $m$ the azimuthal order of the mode
and $\beta$ the Ledoux constant \cite{ledoux1951}.

Uytterhoeven et al. \cite{uytterhoevenmathiasporetti2008} identified some
of the observed modes of HD~49434 as $\ell=$3--8 prograde modes, and estimated the value of
the equatorial velocity to be $v_{\rm eq}=236$~km~s$^{-1}$.  Even if we adopt as equatorial velocity
$v\,\sin\,i=87$~km~s$^{-1}$, the $\ell=6$ modes split by rotation can reach values of the order of the highest observed frequency (12
c/d). Therefore,  the observed frequencies can be explained either by a combination
of $\gamma$ Dor and $\delta$ Sct type modes, or by the splitting of high degree
$g$-modes. Present observations do not allow us to confirm the hybrid nature of HD~49434.

Handler \cite{handler2009} performed a frequency analysis for HD~8801 using
ground-based (GB) photometry. No mode identification is available but
due to the limitations of GB photometry,~we~chose~to~restrict~our study 
to $\ell~\leq~3$ with $v_{\rm eq}=v\,\sin\,i=55$~km~s$^{-1}$ \cite{henryfekel2005}. In~this
case, split $g$-modes are not sufficient to explain
the highest observed frequencies. Therefore the spectrum of HD~8801 can most probably be attributed to hybrid pulsations.

\section{Conclusion}

Using non-adiabatic computations including TDC treatment for models with $\alpha_{\rm MLT} = 2.0$, we predict the excitation of both $\gamma$~Dor 
and $\delta$~Sct modes separated by a region of stable modes in models located in the region of the HR diagram 
where hybrid candidates have been detected.
Moreover, from a comparison between theoretical excited frequencies including the first order effect of
rotation and observed frequencies of HD~49434 and HD~8801, we emphasize that it is necessary to consider 
the effect of rotation on PFs case by case in order to characterize these candidates as hybrid pulsators 
or as $\gamma$~Dor stars with $g$-modes split by rotation.

\IfFileExists{\jobname.bbl}{}
 {\typeout{}
  \typeout{******************************************}
  \typeout{** Please run "bibtex \jobname" to optain}
  \typeout{** the bibliography and then re-run LaTeX}
  \typeout{** twice to fix the references!}
  \typeout{******************************************}
  \typeout{}
 }

\end{document}